\documentclass{ws-procs9x6}
\usepackage{bbm,booktabs}

\begin{document}

\title{Hyperon-nucleon interaction and baryonic contact terms
in SU(3) chiral effective field theory}

\author{S.~Petschauer$^*$}

\address{
Physik Department, Technische Universit\"{a}t M\"{u}nchen,\\
D-85747 Garching, Germany\\
$^*$E-mail: stefan.petschauer@ph.tum.de
}

\begin{abstract}
In this proceeding we summarize results for baryonic contact terms derived within SU(3) chiral effective field theory.
The four-baryon contact terms, necessary for the description of the hyperon-nucleon interaction, include SU(3) symmetric and explicit chiral symmetry breaking terms.
They also include four-baryon contact terms involving pseudoscalar mesons, which become important for three-body forces.
Furthermore we derive the leading order six-baryon contact terms in the non-relativistic limit and study their contribution to the $\Lambda NN$ three-body contact interaction.
These results could play an important role in studies of hypernuclei or hyperons in nuclear matter.

\end{abstract}


\bodymatter

\section{Introduction}\label{sec:intro}

The nuclear forces are very well described within the framework of SU(2) chiral effective field theory\cite{%Bernard1995,
Epelbaum2009,Machleidt2011,Holt2013}.
Therefore it is natural to extend this scheme to the strangeness sector and use SU(3) chiral effective field theory to describe the interaction between baryons, as has been done in a recent calculation of the hyperon-nucleon interaction at next-to-leading order\cite{Haidenbauer2013a}. There the unresolved short-distance dynamics is encoded in four-baryon contact terms with {\it a priori\/} unknown low-energy constants.
These contact terms are constructed according to the symmetries of QCD.

It is not only interesting to understand baryon-baryon scattering itself, but these interactions are also input for studies of hypernuclei and hyperons in nuclear matter, such as exotic neutron star matter.
Especially for the few- and many-body calculations, not only two-baryon interactions, but also three-baryon interactions will play an important role.\cite{Bhaduri1967,Gal1971,Lonardoni2013}
To support the recently observed two-solar-mass neutron stars\cite{Demorest2010,Antoniadis2013}, a very stiff equation of state and therefore a repulsive hyperon-nucleon force is needed.
In order to achieve enough repulsion, it might be necessary to include hyperon-nucleon-nucleon three-body forces as well.
As a first step the leading three-baryon contact forces have been derived and classified within SU(3) chiral effective field theory.

\section{Four-baryon contact terms}\label{sec:2bc}

\begin{figure}[t]
 \centering
 \includegraphics[scale=0.5]{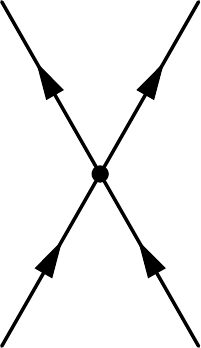}\quad\quad\quad
 \includegraphics[scale=0.5]{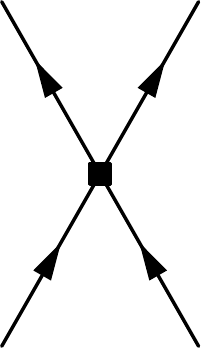}\quad\quad\quad
 \includegraphics[scale=0.5]{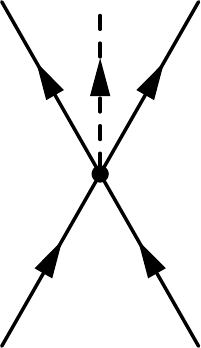}\quad\quad\quad
 \includegraphics[scale=0.5]{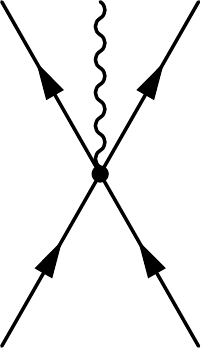}
 \caption{Examples for four-baryon contact vertices at LO and NLO. The dashed and wavy lines denote pseudoscalar mesons and photons, respectively.} \label{fig:ct}
\end{figure}

Considering the baryon-baryon interaction, at leading order one has non-derivative four-baryon contact terms and at next-to-leading order four-baryon contact terms with two derivatives (Fig.~\ref{fig:ct}), both SU(3) symmetric.
Additionally, at next-to-leading order pure baryon-baryon contact terms proportional to quark masses arise and lead to explicit SU(3) symmetry breaking.
Furthermore, at next-to-leading order occur four-baryon contact terms involving one or more pseudoscalar mesons and/or external electroweak fields, cf.~Fig.~\ref{fig:ct}.
These come into play in the description of chiral many-body forces and currents
relevant for few-baryon systems.

By employing the external fields method\cite{Gasser1984,Gasser1985}, we have constructed a complete set of terms for the covariant chiral baryon-baryon contact Lagrangian in flavor SU(3) up to order \(\mathcal{O}(q^2)\)\cite{Petschauer2013a}, which provides the above mentioned contact terms.
The constructed terms are invariant under charge conjugation, parity transformation, time reversal, Hermitian conjugation and local chiral transformations and include Goldstone bosons as well as external fields.
In the case of pure baryon-baryon interaction one obtains a minimal set of 40 terms in the chiral contact Lagrangian up to \(\mathcal O (q^2)\).
After a non-relativistic reduction and a decomposition into partial waves, 28 of these contact terms lead to SU(3) symmetric contributions to the potentials for the channels \({}^1S_0\), \({}^3S_1\), \({}^1P_1\), \({}^3P_0\), \({}^3P_1\), \({}^3P_2\), \({}^3D_1\leftrightarrow{}^3S_1\) and \({}^1P_1\leftrightarrow{}^3P_1\).
Only one specific term leads to an antisymmetric spin-orbit interaction and therefore a spin singlet-triplet mixing.
Such transitions are not possible in the \(NN\) interaction without SU(2) symmetry breaking.
The remaining 12 low-energy constants contribute to the \({}^1S_0\) and \({}^3S_1\) partial waves and lead to SU(3) symmetry breaking contributions linear in the quark masses.

\section{Six-baryon contact terms}\label{sec:3bc}

\begin{figure}[t]
\centering
\includegraphics[scale=.5]{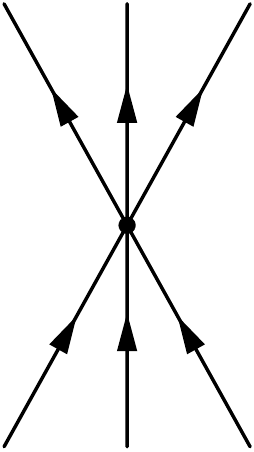}
\caption{Feynman diagram for the leading three-baryon contact force.}
\label{fig:BBBct}
\end{figure}

To describe the effect of three-body forces the full set of the leading three-baryon interactions with a minimal number of low-energy constants, consistent with the symmetries of QCD, is needed.
In contrast to phenomenological approaches,\cite{Bhaduri1967,%Bhaduri1967a,
Gal1971%,Gal1972,Gal1978
} we want to construct these three-baryon forces within the framework of SU(3) chiral effective field theory (following the construction of the chiral nuclear forces\cite{Epelbaum2009}).
We start this construction by deriving the short-range contact contribution in SU(3) chiral effective field theory in Fig.~\ref{fig:BBBct}.

The construction of the chiral Lagrangian works analogously to the construction of the four-baryon contact terms.
The (approximate) symmetries of QCD have to be fulfilled namely charge conjugation, parity transformation, time reversal, Hermitian conjugation and chiral symmetry.
For the construction of the potential it is also important to include the Pauli exclusion principle, since one starts with an overcomplete set of terms in the Lagrangian and wants to obtain the minimal number of low-energy constants. This can be achieved by an antisymmetrization in both the initial and the final state.
The Pauli principle is not as restrictive as for the three-nucleon contact interaction, where one ends up with only one low-energy constant, but it is still relevant.

After constructing the pertinent terms in the chiral Lagrangian and performing a non-relativistic reduction, the leading potentials can be expressed through the following operators in spin space
\begin{equation*} \label{eq:spinbasis}
 \mathbbm1\,,\quad
 \vec\sigma_1\cdot\vec\sigma_2\,,\quad
 \vec\sigma_1\cdot\vec\sigma_3\,,\quad
 \vec\sigma_2\cdot\vec\sigma_3\,,\quad
 \mathrm i\, \vec\sigma_1\times\vec\sigma_2\cdot\vec\sigma_3\,.
\end{equation*}
For each three-baryon channel the prefactors of these spin operators are a combination of SU(3) coefficients and low-energy constants.

As a result we have obtained that the symmetry constraints lead to 18 low-energy constants for the three-baryon contact force.
Table~\ref{tab:3bc} gives the number of additional constants that are introduced with increasing strangeness of the three-baryon system.
\begin{table}[t]
\tbl{Results for the low-energy constants of the leading three-baryon contact terms.\label{tab:3bc}}
{\begin{tabular}{ccccc}\toprule
 strangeness & transitions between & \# constants \\\colrule
 $\phantom{-}0$ & $NNN$ &\phantom{+}1\\
 $-1$ & $\Lambda NN,\Sigma NN$ &+7\\
 $-2$ & $\Lambda\Lambda N,\Lambda\Sigma N,\Sigma\Sigma N, \Xi NN$ & +9\\
 $-3$ & $\Lambda\Lambda\Sigma,\Lambda\Sigma\Sigma,\Sigma\Sigma\Sigma,\Xi\Lambda N,\Xi\Sigma N$ & +1\\
 $-4$ & $\Xi\Xi N, \Xi\Lambda\Lambda,\Xi\Lambda\Sigma,\Xi\Sigma\Sigma$ & +0\\
 $-5$ & $\Xi\Xi\Lambda,\Xi\Xi\Sigma$ & +0\\
 $-6$ & $\Xi\Xi\Xi$ & +0\\\colrule
 & & 18\\\botrule
\end{tabular}}
\end{table}
As an explicit example we give the form of the potentials for the \(\Lambda NN\) interaction with isospin 0 and 1:
\begin{align*}
 V^{I=0}_{\Lambda NN\to\Lambda NN} &= e_2( \mathbbm1+\tfrac13\vec\sigma_2\cdot\vec\sigma_3) +e_3\left( \vec\sigma_1\cdot\vec\sigma_2 +\, \vec\sigma_1\cdot\vec\sigma_3 \right)\,,\\
 V^{I=1}_{\Lambda NN\to\Lambda NN} &= e_4( \mathbbm1 -\, \vec\sigma_2\cdot\vec\sigma_3)\,.
\end{align*}
Note that the low-energy constants for these two isospin channels are independent.
The constant \(e_1\) (proportional to \(c_E\), which is present in the purely nucleonic sector \cite{Epelbaum2009}) does not appear in the \(\Lambda NN\) interaction. Therefore the \(\Lambda NN\) contact interaction can not be constrained by the three-nucleon sector.

\section*{Acknowledgments}
This work has been supported in part by DFG and NSFC (CRC110) and the ``TUM Graduate School''.
I thank J.~Haidenbauer, N.~Kaiser, $\text{U.-G.}$~Mei\ss{}ner, A.~Nogga and W.~Weise
 for fruitful collaboration.


\begin{thebibliography}{10}

\bibitem{Epelbaum2009}
E.~Epelbaum et~al., {\em Rev.Mod.Phys.} {\bf 81}, 1773 (2009).

\bibitem{Machleidt2011}
R.~Machleidt and D.~Entem, {\em Phys.Rept.} {\bf 503}, 1 (2011).

\bibitem{Holt2013}
J.~W. Holt, N.~Kaiser and W.~Weise, {\em Prog.Part.Nucl.Phys.} {\bf 73}, 35
  (2013).

\bibitem{Haidenbauer2013a}
J.~Haidenbauer et~al., {\em Nucl.Phys.} {\bf A915}, 24 (2013).

\bibitem{Bhaduri1967}
R.~Bhaduri, B.~Loiseau and Y.~Nogami, {\em Annals Phys.} {\bf 44}, 57 (1967).

\bibitem{Gal1971}
A.~Gal, J.~Soper and R.~Dalitz, {\em Annals Phys.} {\bf 63}, 53 (1971).

\bibitem{Lonardoni2013}
D.~Lonardoni, S.~Gandolfi and F.~Pederiva, {\em Phys.Rev.} {\bf C87},
  041303 (2013).

\bibitem{Demorest2010}
P.~B. Demorest et~al., {\em Nature} {\bf 467}, 1081 (2010).

\bibitem{Antoniadis2013}
J.~Antoniadis et~al., {\em Science} {\bf 340}, 6131 (2013).

\bibitem{Gasser1984}
J.~Gasser and H.~Leutwyler, {\em Annals Phys.} {\bf 158}, 142 (1984).

\bibitem{Gasser1985}
J.~Gasser and H.~Leutwyler, {\em Nucl.Phys.} {\bf B250}, 465 (1985).

\bibitem{Petschauer2013a}
S.~Petschauer and N.~Kaiser, {\em Nucl.Phys.} {\bf A916}, 1 (2013).

\end{thebibliography}
\end{document}